# A Giant, High-frequency Oscillation of Metal Strip Actuator under DC Electric Field

Chengyi Xu, Fanyi Cai, Baozhang Li, Chunye Xu, Jianming Zheng[*]

Hefei National Laboratory for Physical Sciences at the Microscale, Department of Polymer Science and Engineering, University of Science and Technology of China, Hefei 230026, P.R. China

**ABSTRACT**

Different from conventional electroactive polymers, here we firstly present a new facile actuator made from aluminum alloy. The high-frequency electrically induced flapping motion was characterized under varied physical factors. This electroactuation results from alternative processes of charge induction and discharge, which is confirmed by the existence of periodical pulse current in the circuit. The metal actuator is of great stability and can maintain several days if not for any structural fatigue. Easy fabrication, high tunable frequency and durability make it potential for implementation of actuators for sensors, microelectromechanical systems and robotics.

**Keywords**: actuator, flapping-wing motion, electroactuation, metal strip

## 1. INTRODUCTION

Smart materials are known for their excellent properties in response to external stimuli, such as electric field[1], light[2], pH[3] or temperature[4]. Within the past two decades, electroactive polymers (EAPs) including ferroelectric polymers, conductive polymers and dielectric elastomers, have emerged as a new class of smart materials for the realization of artificial muscles, robotics and adaptive optical electronics[5]. These sandwich-structured EAP actuators have attracted much attention for their remarkable abilities to convert electrical energy into mechanical work by reversibly changing volume, shape or bending behavior. Recently, carbon nanotubes[6] and their functionalized derivatives[7] have been designed to improve actuation performance as dopants and electrode materials, taking advantage of unique characteristics as high electrical conductivity and large specific surface area. However, inconvenient fabrication, solvent participation, low mechanical output density and short lifetime still largely restricted the commercial applications of conventional EAPs. Different from previous research, here we present a new facile tunable metal actuator with high-frequency flapping-wing oscillation under DC electric field.

## 2. EXPERIMENT

In the current study metal strip samples (thickness 11.6 μm) were prepared from flat aluminum alloy sheet (CEO-465, China). Two pieces of square indium tin oxide coated glass served as a pair of electrodes with the gap distance kept 4.0 cm. The conductive layers were inward face-to-face so that parallel electric field could be formed in the internal space. Insulating supporters made from silicone rubber (Xuele Reagents, China) were shaped with a horizontal crack to clamp the sample. Then we positioned the supporter into the center of electric field and the strip was kept flat as possible. The actuation of the clamped strip was generated by a high voltage power source (Model DW-P503, Tianjin Dongwen, China), providing DC voltage up to 25 kV. The whole electrically induced motions

---

[*] E-mail: jmz@ustc.edu.cn; Tel &Fax: +86-551-360-3470; http://www.hfnl.ustc.edu.cn/2009/0605/955.html
Address: Hefei National Laboratory for Physical Sciences at the Microscale, University of Science and Technology of China, Hefei, 230026, P. R. China

were recorded by a high speed camera (Fast Cam SAX, Photron, USA) at a speed of 250 frames per second for further research.

## 3. RESULTS AND DISCUSSION

One cycle of flapping-wing oscillation with large displacement is demonstrated in Fig. 1(a) (also see supporting video). The metal strip reversibly bends upward and downward between two electrodes at a high frequency.

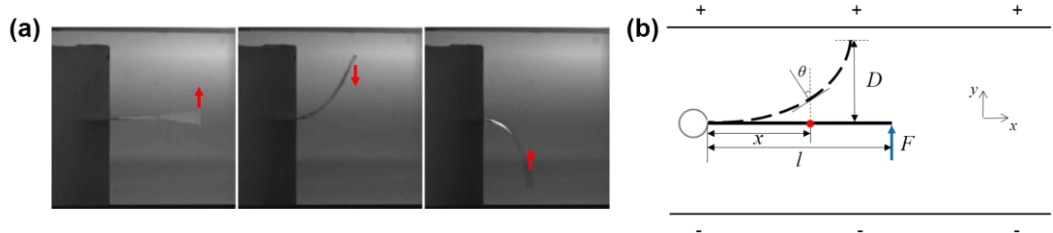

**Fig. 1.** (a) Images of realtime flapping-wing motion recorded by high speed camera; and (b) schematic of strip actuator for electrical and mechanical analysis.

Applied voltage is found to be linearly proportional to flapping frequency. Physical factors covering a wide range of properties were taken into our consideration. Dependence of frequency directly correlates with the strip size. For both the increase of width and the decrease of length, the strip actuator requires higher voltage to overcome the extra equivalent bending torque as Fig. 2(a) and (b). Among three tip shapes of actuators tested, Fig. 2(c) reveals that sharper tip leads to higher frequency. Such a difference is believed to be caused by the fact that charges are more likely to accumulate at large curvature, which benefits in accelerating the point discharge process. The oscillation frequency reaches a maximum of 48 Hz at 20 kV with a triangle shape, comparable to those of hummingbirds. Furthermore, considering the uneven distribution of actual electric field, we positioned the actuator in order every centimeter at two different lines as plotted. It turns out that the flapping frequency is relatively stable in the center area and slightly declines when it comes to the margin as shown in Fig. 2(d), suggesting that the field intensity weakens with respect to the reducing distance from the center.

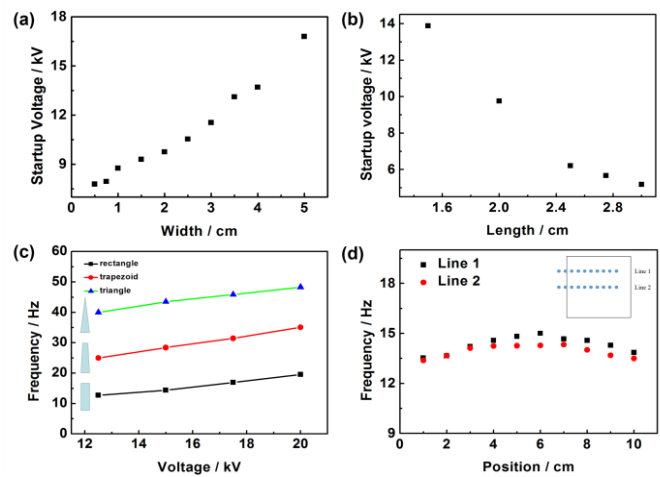

**Fig. 2.** Effects of physical factors on the oscillation of strip actuator. Dependence of actuators with (a) varied widths from 0.5 cm, 0.75 cm, 1 cm, 1.5 cm, 2 cm, 2.5 cm, 3 cm, 3.5 cm, 4 cm to 5 cm; (b) varied lengths from 1.5 cm, 2.0 cm, 2.5 cm, 2.75 cm to 3 cm; (c) different tip shapes as rectangle, trapezoid and triangle; and (d) positions in the electric field.

The origin of oscillation is derived from a reciprocating charge transfer process. After a DC potential is applied, an electrostatic field instantly forms between the two electrodes. Under an ideal condition, a piece of metal strip perpendicular to uniform electric field keeps still due to an equilibrium state of force. However, any bias in practice, such as the variation of DC electric field and the roughness of strip surface, could break this balance. Electrons tend to aggregate at the free end. Subsequently, this part is attracted by a resultant force in the direction of positive electrode. The exerted force is supposed to become larger as the distance to electrode gradually decreases. Consecutive charge redistribution on the conductive strip surface also greatly facilitates this actuation. Then the strip tip comes in contact with the electrode, releasing all electrons. This discharge process usually takes place at the tip owing to the great curvature. Upon completion, the metal strip is immediately injected with opposite charges and thus driven to the counter electrode. To study the realtime locomotion of any point $x$ on the strip surface as illustrated in Fig. 1(b), the immediate state can be divided into two separate parts, one is the surface charge distribution controlled by electrostatic induction and the other is the bending behavior controlled by resultant electrostatic force. The charge distribution and electric potential at $x$ follow the relationship as $U(x) = 1/4\pi k\varepsilon_0 \int \sigma(x)/|d_0 - y| \, dS$, which directly affects the force condition with two dependent equations as $F(x) = \varepsilon_0 U^2/2(d_0 - y)^2$ and $EIy = F(x-l)^3/6 - Fl^2x/2 + Fl^3/6$, where $E$ is the elastic modulus of the strip, $I$ is the moment of inertia, $U$ is the applied voltage, $d_0$ is the initial distance, $F(x)$ is the force distribution at $x$, $S$ is the integral area and $\varepsilon_0$ is the space dielectric constant. In our study, images taken by high speed camera reveal that the whole oscillation of metal beam can even occur without touching the electrodes. It is interpreted that air molecules near the electrodes are partly ionized so that the plasma mixture plays a crucial role as mediate charge carrier in this non-touching circumstance.

During the flapping oscillation, a tiny current forms in the circuit to avoid electric breakdown. We designed our device by connecting a tunable resistance in series and an oscilloscope (RIGOL DS1062CA, China) was applied to record the voltage $V_R$ across the resistance. The current shift was characterized by measuring $V_R$ according to electric principles. In Fig. 3, it is confirmed that pulse current does exist within the reciprocating oscillation, which identifies our interpretation above of air molecules' role as charge carriers. The measured pulse current signal has a steep front edge, which indicates the discharge process takes place in a short time of milliseconds. Each cycle has two separate pulses and the oscillating frequency can be yielded by dividing the pulse number per second by a factor of 2. The obtained frequency matches well with the value acquired from high speed camera. Furthermore, the pulse signal patterns become much denser with the increasing voltage corresponding to higher electro-induced frequency. Though the metal actuator requires a high voltage (>10 kV) to activate, the small current (up to $8 \times 10^{-6}$ A) in the circuit ensures a relatively low energy consumption (about 80 mW). It is also notable that this electroactuation is of great stability and can maintain several days if it were not for any structural fatigue.

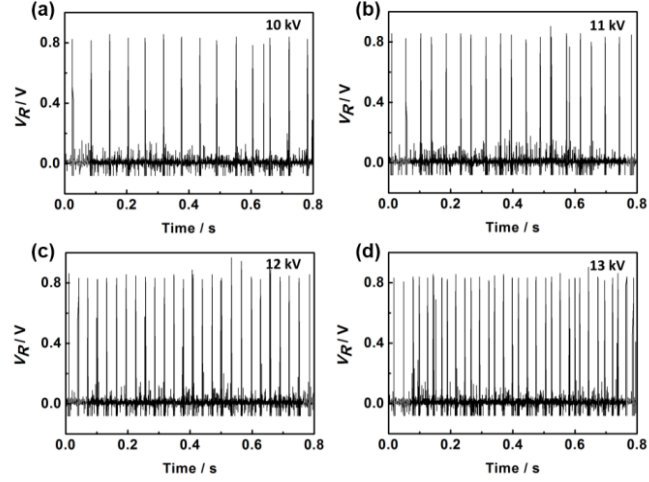

**Fig. 3.** Periodical pulse signals across the resistance (R= 0.1 MΩ) induced by the reciprocating oscillation at the applied voltage of (a) 10 kV (b) 11 kV (c) 12 kV and (d) 13 kV.

To further characterize our prototype actuator, instantaneous vertical displacement (*D*) is measured as the free tip distance deviated from the initial position. After tracking massive periodic motions, it is found that the curve of strip tip locomotion agrees well with sine wave equation, behaving like a periodical harmonic oscillator (Fig. 4(a)). We also observed that if the external voltage alternately increases and decreases as a load-unload cycle, a dynamic lag appears and the frequency tends to form a hysteresis loop as in Fig. 4(b). To start bending, the actuator requires higher voltage to overcome the initial energy barrier ($V_s$), which appears to be several kilovolts higher than the minimum voltage to maintain oscillation ($V_m$). Then in the unloading process when the input voltage reversibly decreases from 17.5 kV, lower voltage is required to maintain flapping motion since the oscillating strip already has its extra kinetic energy.

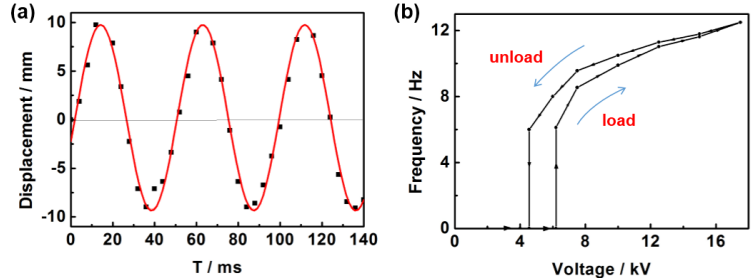

**Fig. 4.** (a) Curves of tip displacement with respect to time, and (b) hysteresis loop found within one load-unload cycle.

For prospect applications, we can simplify the strip actuator into a cantilever model, which can be applied to electrically activated balance. The fundamental frequency can be obtained according to $f = \sqrt{k/m}/(2\pi)$, where *m* is the equivalent mass of the cantilever. The variable *k* is the effective spring constant and is given as $k = 3EI/L^3$, where *E* is Young's modulus, *L* is the length, and *I* is the moment of inertia in the bending motion. The fundamental frequency of the strip can be determined by *m* at a given voltage from the equation. After testing the electrically induced resonance frequency, the value of attached mass weight at the cantilever tip can be acquired.

## 4. CONCLUSION

We demonstrated a high-frequency flapping-wing motion of metal actuator under DC electric field. The oscillation was recorded by a high speed camera and characterized by analyzing instant state of movement. Different from that of conventional polymer actuators in mechanism, this electroactuation is attributed to alternative processes of charge induction and discharge, which can be confirmed by the existence of periodical pulse current in the circuit. A hysteresis loop is found for strip actuator to overcome the initial energy barrier. Applied voltage is nearly proportional to oscillation frequency. This electrically induced oscillation depends on several physical factors including strip size, tip shape as well as the field position. Easy fabrication, high tunable frequency and durability make it potential for another implementation of actuators for sensors, microelectromechanical systems and biomimetics.

## ACKNOWLEDGEMENTS

This work received financial support from the National Natural Science Foundation of China (21273207, 21074125), Anhui Province Natural Science Foundation, China (11040606M57), the National Basic Research Program of China (2010CB934700), the "Hundred Talents Program" of CAS, and the "National Thousand Talents Program".